\documentclass[apj]{emulateapj}
\usepackage{color}
\usepackage{graphicx}

\begin{document}

\title{Radio transient following FRB 150418: afterglow or coincident AGN flare?}

\author{Ye Li \& Bing Zhang}

\affil{Department of Physics and Astronomy, University of Nevada, Las Vegas, NV 89154, USA}

\begin{abstract}

Recently, Keane et al. reported the discovery of a fading radio transient following FRB 150418, and interpreted it as the afterglow of the FRB. Williams \& Berger, on the other hand, suggested that the radio transient is analogous to a group of variable radio sources, so that it could be a coincident AGN flare in the observational beam of the FRB. A new observation with VLA showed a re-brightening, which is consistent with the AGN picture.
Here, using the radio survey data of Ofek et al., we statistically examine the chance coincidence probability to produce an event like the FRB 150418 transient. We find that the probabilities to produce a variable radio transient with at least the same variability amplitude and signal-to-noise ratio as the FRB 150415 transient, without and with the VLA point, are $P_1 \sim 6 \times 10^{-4}$ and $P_1 \sim 2 \times 10^{-3}$, respectively. In addition, the chance probability to have a fading transient detected following a random time (FRB time) is less than $P_2 \sim 10^{-2.9\pm 1.3}$. Putting these together and assuming that the number of radio sources within one Parkes beam is 16, the final chance coincidence of having an FRB 150418-like radio transient to be unrelated to the FRB is $< 10^{-4.9\pm1.3}$ and $< 10^{-4.4\pm1.3}$, respectively, without and with the VLA point. We conclude that the radio transient following FRB 150418 has a low probability being an unrelated AGN flare, and the possibility of being the afterglow of FRB 150418 is not ruled out.

\end{abstract}

\section{Introduction}
Fast Radio Bursts (FRBs) are bright milliseconds radio transients
with dispersion measure (DM)
much larger than the values expected for the Milky Way galaxy,
so that they are expected to have an extragalactic origin
\citep{2007Sci...318..777L, 2011MNRAS.415.3065K, 2013Sci...341...53T, 
2014ApJ...790..101S, 2014ApJ...792...19B, 2015arXiv151107746C, 2015Natur.528..523M, 
2015ApJ...799L...5R,2015MNRAS.447..246P,2016Natur.530..453K}.
However, due to the large positional uncertainty ($\sim$ 14 arcminutes), no FRB has been previously well located to allow a distance measurement.

Recently, \cite{2016Natur.530..453K} reported the discovery of a radio transient following FRB 150418 starting from 2 hours after the FRB, which faded away in $6-8$ days. They claimed that the radio transient is the afterglow of the FRB, and used it to identify an elliptical galaxy at $z = 0.492 \pm 0.002$. The measured baryon number density $\Omega_b$ based on this redshift is consistent with the value inferred from the CMB data, lending an indirect support to the association. If such an association is established, the brightness of the radio afterglow is consistent with that of a double compact star merger system (\cite{2016arXiv160208086Z} and references therein).

However, shortly after the publication of the discovery, \cite{2016arXiv160208434W} suggested that the chance probability of having a radio variable source within one Parkes beam is of the order of unity, and argued that the so-called afterglow of FRB 150418 discovered by \cite{2016Natur.530..453K} could be simply an unrelated AGN radio flare in spatial coincidence with the FRB. \cite{2016ATel.8752....1W} further observed the source and claimed a re-brightening of the host, which strengthened the AGN interpretation.

Here we statistically examine the chance probability of having a transient source similar to the putative afterglow of FRB 150418. We notice the following important facts: 1. The first two data points in 5.5 GHz of \cite{2016Natur.530..453K} are 0.27(5) mJy per beam and 0.23(2) mJy per beam, respectively, which are a factor of 2.5-3 times of the claimed ``quiescent" flux 0.09 mJy per beam at later time, as shown in the left panel of Figure \ref{simul}. The standard deviation is too large for a flux fluctuation due to scintillation, suggesting that these two points have to be related to a flaring object.  2. The flux is fading during the first 6-8 days after the FRB. Our simulations intend to address the probability of having a variable source with such a large flux variability in the FRB observation beam (Sect. 2), and the probability to have a bright flare that is fading from 2 hours to 8 days after the FRB (Sect. 3). We find a very small overall probability. We also compare some short GRB radio afterglows with the FRB 150418 transient, and find that the probabilities that some short GRB afterglows to be confused as AGN flares due to chance coincidence could be comparable to that of the putative afterglow of FRB 150418 (Sect. 4). Putting it together, we argue that the possibility that the radio transient is the intrinsic afterglow of the FRB is not ruled out.

\section{Spatial and flux coincidence probability}

\cite{2016arXiv160208434W} argued that the probability of finding a radio source in each Parkes beam is $N_{\rm r}=16$, 
and that of finding a variable radio source is $\sim 0.6$. However, the true question is: what is the probability of finding a variable radio source with the amplitude and S/N at least those of the transient following FRB 150418? 

\begin{figure*}[!htb]
\centering
\includegraphics[width=1.0\columnwidth]{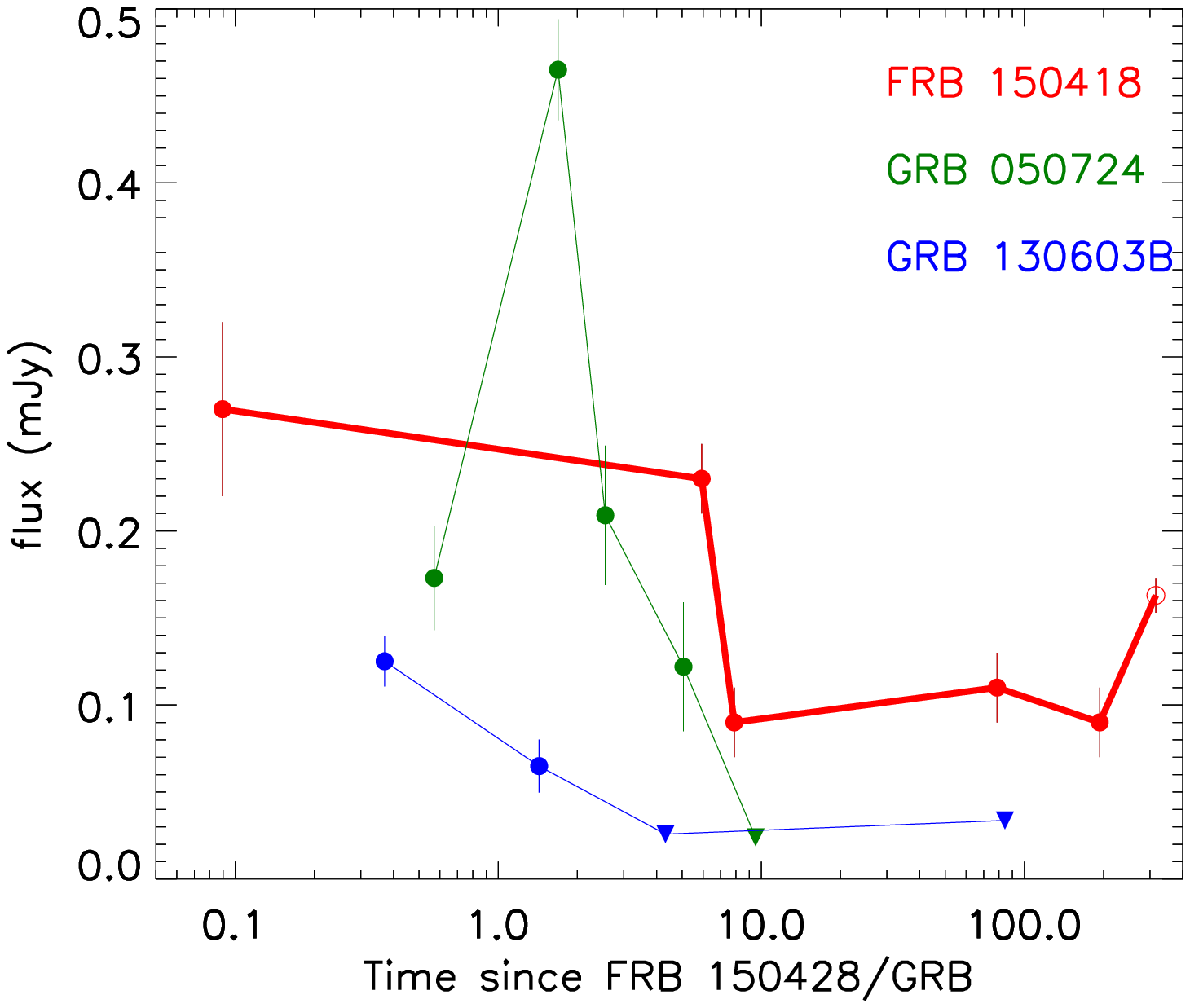}
\includegraphics[width=1.0\columnwidth]{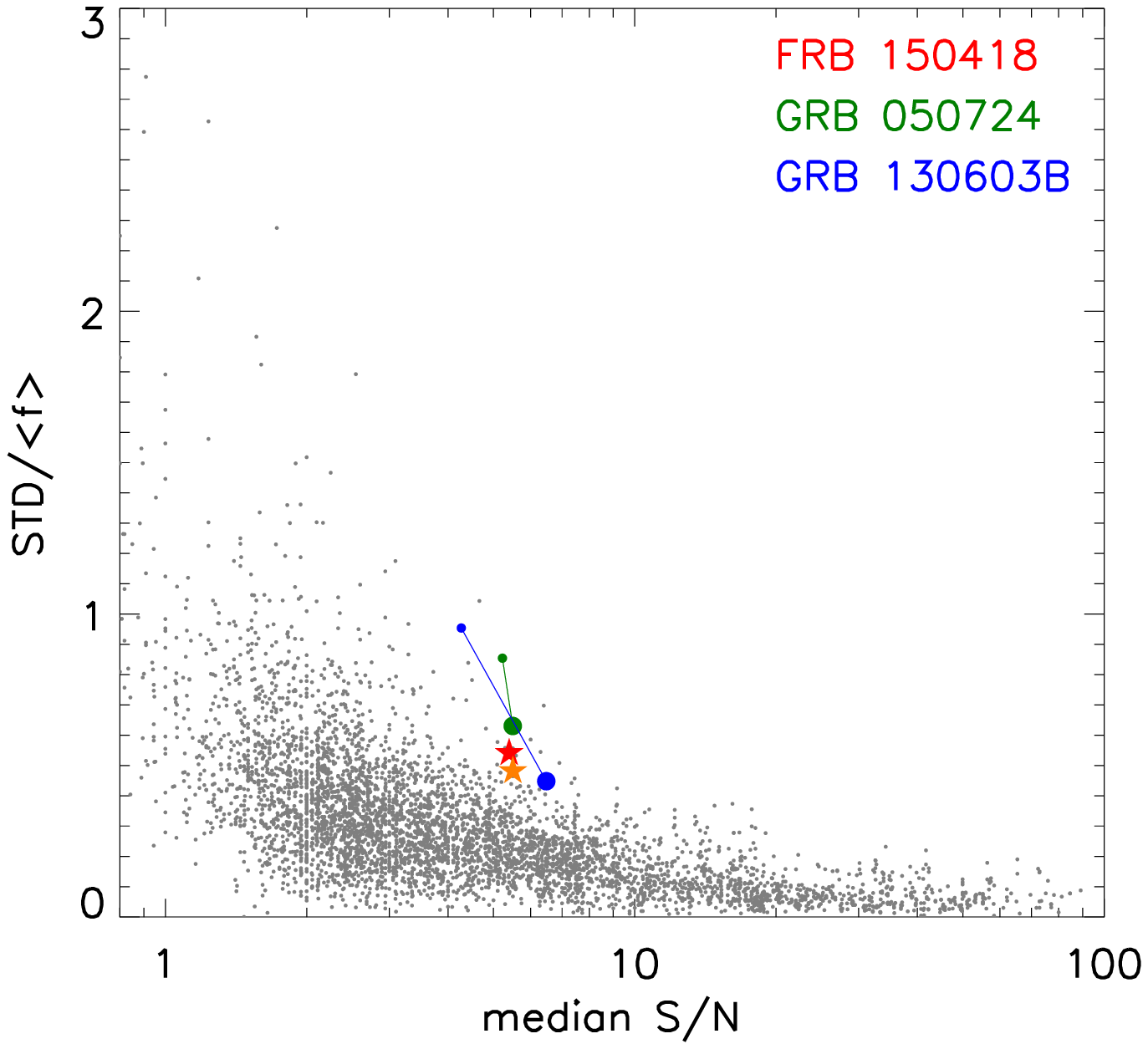}
\caption{
Left: The putative radio afterglow lightcurve of FRB 150428 (red). 
The last point is from the latest VLA point \citep{2016ATel.8752....1W}, 
while the first five points are original ATCA points \citep{2016Natur.530..453K}.
The radio afterglow light curves of the short GRBs 050724 (green) and 130603B (blue)
are also presented for comparison.
Right: The relative standard deviation STD/$\langle f \rangle$ vs.
median S/N for our simulated mock observations (gray dots) compared with
the FRB 150418 afterglow candidate (stars) and two short GRBs (filled circles).
The red star denotes the original ATCA data \citep{2016Natur.530..453K},
and the orange star also includes the latest VLA data \citep{2016ATel.8752....1W}.
The short GRBs 050724 (green) and 130603B (blue) are also shown.
For each short GRB, the small circles treat upper limits as detections assuming that
the fluxes are half of the upper limits and the errors are also half of the upper limits.
The large circles, on the other hand, exclude the upper limits from the calculations.
The true value is likely between the two circles, as shown as the green and blue segments.
}
\label{simul}
\end{figure*}

To address this question, similar to  \cite{2016arXiv160208434W}, we make use of the 5 GHz survey data with Very Large Array (VLA) and the expanded VLA by \cite{2011ApJ...740...65O}.
There are 464 sources in the main catalog of \cite{2011ApJ...740...65O}.
For each source, there are $14-16$ observations (see Table 6 of \cite{2011ApJ...740...65O}, which lists all the observation results). 
In order to compare with the observation of FRB 150418 afterglow candidate \citep{2016Natur.530..453K},
we randomly select the same number of observational epochs as FRB 150418 for each source, 
and calculate the relative standard deviation STD/$\langle f \rangle$
as well as median signal-to-noise ratio S/N.
We do it 1000 times for each source, so that we have 464,000 simulated mock observations.
Similar to Fig. 2 of \cite{2016arXiv160208434W}, 
we present the S/N - STD/$\langle f \rangle$ two-dimensional distribution of these mock events 
in the right panel of Figure \ref{simul}.
For the sake of clear presentation, only a random set of 4,640 mock events are shown.
For comparison, the FRB afterglow candidate, which has STD/$\langle f \rangle = 0.54$ and median S/N $=5.4$ with the 
observational data of \cite{2016Natur.530..453K}, is also marked as the red star in the right panel of Fig. \ref{simul}. 
\cite{2016ATel.8752....1W} reported an observation of the FRB host on 2016 Feb 27 and 28, 
which has a flux 0.157 $\pm$ 0.006 mJy/beam. 
By including this point, the FRB afterglow candidate has
STD/$\langle f \rangle = 0.48$ and median S/N $=5.5$. 
It is shown as the orange star in the right panel of Fig. \ref{simul}.

An immediate observation from Fig.1 is that a large STD/$\langle f \rangle$ tends to appear for small S/N values. 
At the S/N for FRB 150418, the observed STD/$\langle f \rangle$ in general is much smaller than that of the FRB 150418 transient.
\cite{2016arXiv160208434W} argued that the transient source is consistent with the distribution of the \cite{2011ApJ...740...65O} sources.
However, we argue that it is more important to check the chance probability to have a variable source with both  STD/$\langle f \rangle$ 
and S/N at least the values inferred from the FRB 150418 transient. 
From our mock sample, the fraction of events that have 
STD/$\langle f \rangle$ larger than STD/$\langle f \rangle_{\rm FRB}$ 
and median S/N larger than S/N$_{\rm FRB}$ for the \cite{2016Natur.530..453K} data only (without the late VLA point of \cite{2016ATel.8752....1W})
turns out to be $P_1 \sim 6 \times 10^{-4}$. 
It indicates that the average number of events 
which could be as bright as the FRB transient in the Parkes beam by chance
is $N_{\rm v}=N_{\rm r}P_1 \sim 0.009$.
Adding the latest VLA observational point \cite{2016ATel.8752....1W}, this fraction is increased to $P_1 \sim 2 \times 10^{-3}$, and the variable number becomes $N_{\rm v}=N_{\rm r}P_1 \sim 0.03$\footnote{Since the last data point was detected by VLA while other data points were detected with ATCA, there might be some systematic errors introduced by cross-calibration between different telescopes.}.
Therefore, even if the radio variable source may be common, the chance probability of having a high-variability radio transient similar to the putative FRB 150418 afterglow 
within the FRB 150418 Parkes beam is small. 

We also try to compare the FRB afterglow candidate with \cite{2016ApJ...818..105M},
who monitored a larger sky area and have more radio sources in their catalog.
Since there are only two observational points in week timescale for each source in \cite{2016ApJ...818..105M},
we choose the first observational point of \cite{2016Natur.530..453K}, i.e. $0.27\pm0.05$ mJy, and the quiescent flux, $0.09\pm0.02$ mJy,
to compare with those in  \cite{2016ApJ...818..105M}.
Using the distribution of $m=\frac{\Delta f}{\bar f}$, analogous to STD/$\langle f \rangle_{\rm FRB}$, 
and $V_{\rm s}=\frac{\Delta f}{\sigma}$, analogous to median S/N, 
as shown in Figure 10 of \cite{2016ApJ...818..105M}, we find that
there is no source in \cite{2016ApJ...818..105M} that is as significant as 
this FRB afterglow candidate.
If we change the second point to $0.11\pm0.02$,
the fraction of sources as significant as the FRB afterglow candidate is 0.001.
This is consistent with our previous result.

\section{Temporal coincidence probability}

The existence of an event with a similar STD/$\langle f \rangle$ and median S/N to the FRB 150418-transient does not necessarily interpret the observation. 
One important, intriguing fact is that the radio source was fading during the span of the 5 ATCA observations (left panel of Fig.\ref{simul}), which have observational epochs at
$t_0+[0.09, 5.9, 7.9, 78.7, 193.4]$ days after the FRB time $t_0$, respectively, for the \cite{2016Natur.530..453K} observation. Assuming that during 190 days there
was only one bright transient (i.e. there is no variability during the last three observational epoch), the duty cycle of flares may be estimated as
 $\sim$ 8 days /190 days $ \sim 0.04$. 
the chance of having the first observation to be the 
brightest point is $P_2 \sim 0.09/190 \sim 5 \times 10^{-4}$. 

\begin{figure*}[!htb]
\centering
\plottwo{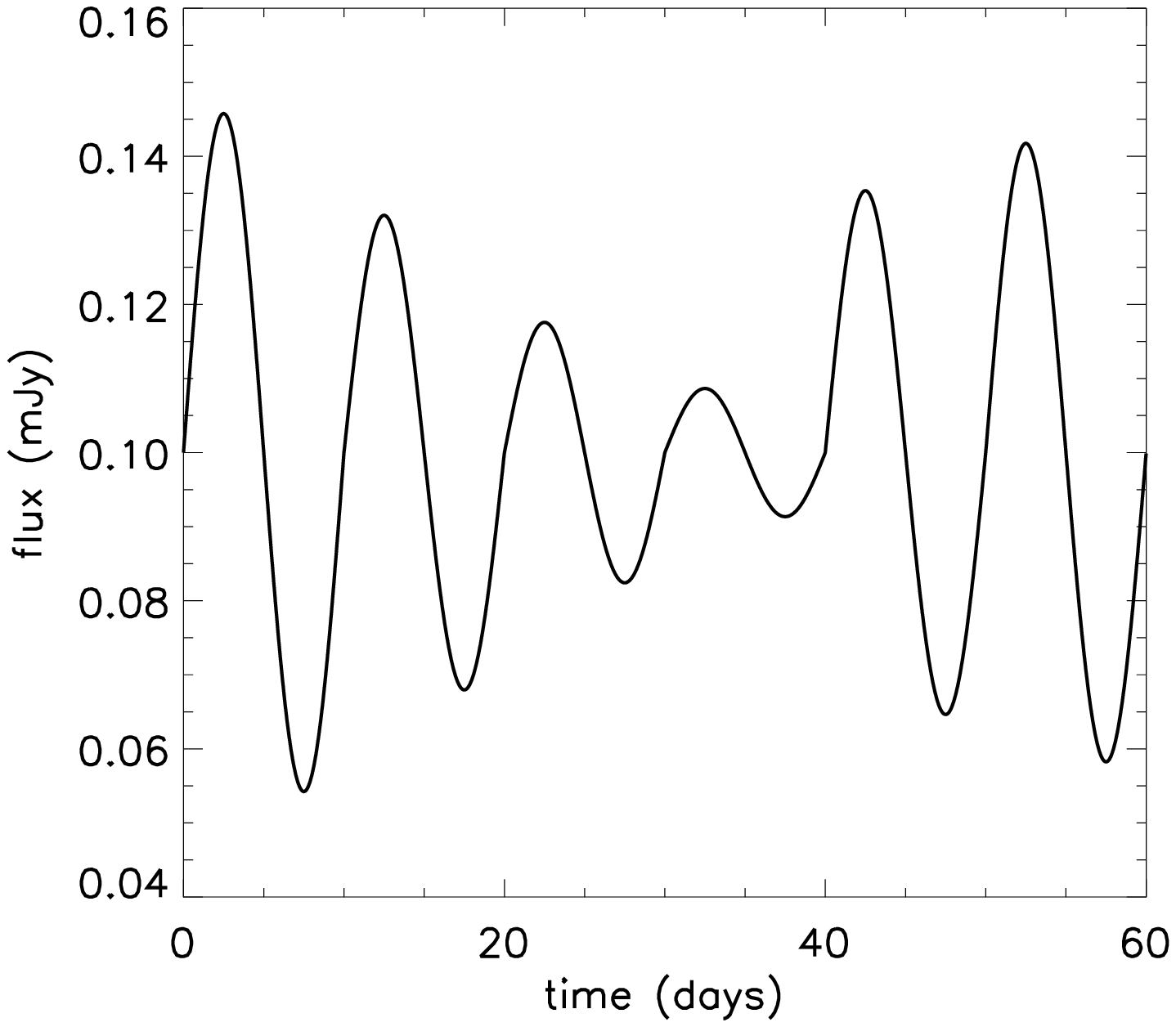}{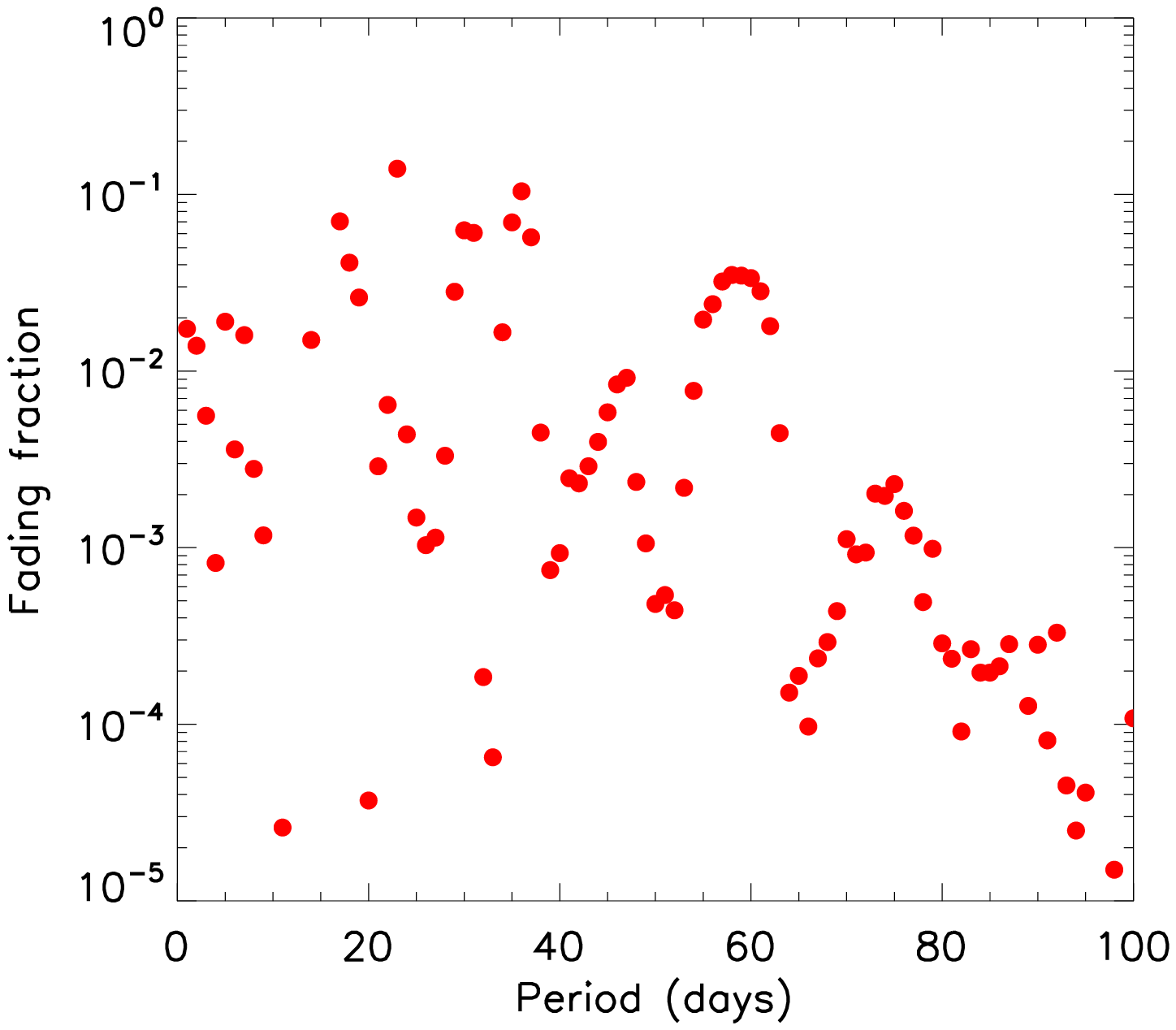}
\caption{
Left: A passage of one simulated lightcurve.
Right: The fraction of simulated detection series similar to
FRB afterglows (monotonously decreasing) as a function of the assumed period.}
\label{tempf}
\end{figure*}

One may argue that the source may be variable all the time, and that the first observation may have missed even brighter phases at earlier times. 
In order to examine these probabilities, we perform a Monte Carlo simulation using the {\em most conservative} approach by assuming that the
source has a variability duty cycle of 100\%\footnote{By introducing a smaller duty cycle, our simulations suggest that the chance probabilities for the
temporal coincidence are indeed even lower.}. We assume that the source
varies sinusoidally with time, i.e.
$$f= f_0+A {\rm sin}(2 \pi \frac{t}{\cal{P}}),$$
but with the amplitude $A$ varying in different periods. Here $f_0$ is the flux of the quiescent state.
Since the period $\cal P$ is unknown, we vary it from 2 days to 100 days.
By fixing a particular $\cal P$, we simulate a mock light curve which lasts for 500 periods.
We allow the amplitude $A$ to be variable. For each period, it is randomly simulated based on the STD/$\langle f \rangle$
distribution in the \cite{2011ApJ...740...65O} catalog.
For a sinusoidal distribution, one has $A=\sqrt{2}\rm STD$, and $\langle f \rangle = f_0$.
In order to be consistent with the observed FRB afterglow, the STD/$\langle f \rangle$ distribution with median S/N $=(5-6)$ is used.
An example of a small passage of the simulated light curve is shown in the left panel of Figure \ref{tempf}.

From each simulated light curve, we randomly pick a $t_0$ time as the epoch of the FRB, and then pick up the fluxes at the 
time series $t_0+[0.09, 5.9, 7.9, 78.7, 193.4]$ days as a simulated detection series. 
We require that the resulting light curve should statistically decrease with time with respect to the first point. 
We simulate 10$^6$ FRBs in each light curve
and estimate the fraction of simulated detection series that satisfy the above criterion.
The fraction as a function of the assumed period is shown in the right panel of Figure \ref{tempf}.
It can be seen that, the fraction of the simulated detection series that satisfy the monotonic criterion $P_2$ is period dependent.
In general, the larger the period is, the less possible to produce a detection series similar to the FRB 150418 transient.
By accounting for the range of $P_2$ introduced by the period-dependence, we finally get $P_{2}=10^{-2.9 \pm 1.3}$, with the
latest probability $P_{2, \rm max} \sim 0.14$ (corresponding to period $\cal{P}$ $\sim 23$ days).

In reality the variable source is not strictly periodic. We also tried to simulate light curves with a distribution of period within each light curve.
Both a uniform distribution and a Gaussian distribution of $\mathcal{P}$ are tested. The mean value of $P_2$ is slightly larger, but the
scatter becomes smaller, with the largest probability $P_{2, \rm max} \sim 0.01$. 

Combining all the constraints (spatial, flux, and temporal coincidence), the final chance probability to have an unrelated AGN flare
to mimic the putative afterglow of FRB 150418 is
$$ P = N_{\rm r} P_1 P_2,$$
which is $\sim 10^{-4.9 \pm 1.3}$ for the original \cite{2016Natur.530..453K} data, and is
$\sim 10^{-4.4 \pm 1.3}$ with the inclusion of the latest VLA data \citep{2016ATel.8752....1W} (see Table \ref{tbP}).
Both are very small numbers.


\begin{center}
\begin{table*}[!htb]
\caption{Probability to reproduce FRB/SGRB radio afterglows by a chance coincidence \label{tbP}}
\begin{tabular}{ccc|c|c|c|cccc}

\hline
name & STD/$\langle f \rangle$ & median S/N & $N_{\rm r}$ & $P_1$ & $P_2$ & $P=N_{\rm r}P_1P_2$ \\
\hline
FRB 150418 (\cite{2016Natur.530..453K}) & 0.54 & 5.4 & 16 & $6 \times 10^{-4}$ & $10^{-2.9\pm1.3}$ & $10^{-4.9\pm1.3}$\\
FRB 150418 & 0.48 & 5.5 & 16 & $2 \times 10^{-3}$ & $10^{-2.9\pm1.3}$ & $10^{-4.4\pm1.3}$ \\
(\cite{2016Natur.530..453K}$+$\cite{2016ATel.8752....1W}) & & & & & &\\
\hline
GRB 050724 (with upper limits) & 0.85 & 5.2 & & $4.3 \times 10^{-5}$ & & \\
GRB 050724 (without upper limits) & 0.63 & 5.5 & & $2 \times 10^{-4}$ & & \\
GRB 130603B (with upper limits) & 0.95 & 4.3 & & $1 \times 10^{-4}$ & &  \\
GRB 130603B (without upper limits) & 0.45 & 6.5 & & $7.8 \times 10^{-4}$ & & \\
\hline
\end{tabular}
\tablecomments{
Column 1:  FRB/GRB names; Column 2: relative standard deviation. STD is the standatd deviation and $\langle f \rangle$ is the average flux.
Column 3: median signal to noise ratio; Column 4: estimated number of radio sources within one Parkes beam;
Column 5: probability to reproduce one event as significant as the FRB afterglow candidate (with both STD/$\langle f \rangle$ 
and median S/N larger than those of the FRB afterglow candidate) with the observational data of \cite{2011ApJ...740...65O}; 
Column 6: probability upper limit to have a fading event by chance, with a 100\% duty cycle assumed;
Column 7: overall probability to have the FRB afterglow candidate originating from chance coincidence.}
\end{table*}
\end{center}

\section{Comparison with short GRB radio afterglows}

Since the putative afterglow of FRB 150418 has a flux comparable to that of short GRBs \citep{2016Natur.530..453K, 2016arXiv160208086Z}, 
it is worth comparing the variability properties of short GRB afterglows with the FRB 150418 transient. 
There are two SGRBs with at least two radio afterglow detections. 
One is GRB 050724 from the radio afterglow catalog of Chandra \& Frail (2012). 
Another one is GRB 130603B, which has two detections and two upper limits
\citep{2014ApJ...780..118F}.
Following the same procedure, we present their STD/$\langle f \rangle$ and median S/N 
in Table \ref{tbP}. Assuming that they fall into one of the Parkes beams, we 
calculate the chance probability of confusing them with an underlying AGN radio flaring sources in the 
 \cite{2011ApJ...740...65O} catalog. 
The afterglow light curves of the two short GRBs are presented along with that of FRB 150418 in the left panel of Fig.\ref{simul}  
(in green and blue). Upside down triangles indicate upper limits.    
In the process of estimating STD/$\langle f \rangle$ and median S/N for the two short GRBs, we treat the upper limits
in two different methods: one is to include them by assuming that both detection values and the errors are half of the upper limit values;
the other is to exclude the upper limits completely. These two methods define a range of STD/$\langle f \rangle$ and median S/N,
which are marked in the right panel of Fig.\ref{simul} as segments connected with two filled circles.
The results with the first method are marked with small circles,
and  those with the second method are marked with large circles.
One can see that GRB 050724 has a more significant variability and a smaller chance probability than FRB 150418.
On the other hand, GRB 130603B is less variable and has a similar chance probability to be confused as a flaring source 
as FRB 150418.
In general, FRB 150418 sits near the two short GRBs in the STD/$\langle f \rangle$ - S/N space, suggesting that
the data are not inconsistent with being an FRB / short GRB afterglow,
as suggested by \cite{2016Natur.530..453K}.

\section{Conclusions and Discussion}

We have statistically examined the probability of having a random variable source, such as AGN, following FRB 150418 by comparing
the event with the radio variable sources presented in \cite{2011ApJ...740...65O}.
By requiring that the coincident transient should have at least the same  STD/$\langle f \rangle$ and median S/N values as 
the FRB 150418 transient and that it should decay starting from the first observation, the combined spatial, flux, and temporal 
chance coincidence probability is $< 10^{-4.9\pm1.3}$ for \cite{2016Natur.530..453K} observational data only,
and $< 10^{-4.4\pm 1.3}$  with the VLA point included \citep{2016ATel.8752....1W}.
We also show that the event is not inconsistent with a short GRB radio afterglow (with host
contamination). We therefore argue that the event has a low probability being an unrelated AGN flare, and that
the possibility that the source is the intrinsic afterglow of FRB 150418 is not ruled out by the current data.
Further monitoring the source is needed to place more constraints on the afterglow and AGN possibilities.
If the source later re-brightens to the level of the first two data points of \cite{2016Natur.530..453K}, then the afterglow
scenario would be ruled out. A smaller variability as seen by \cite{2016ATel.8752....1W}, would not significantly
alter the conclusion of this paper, since scintillations \citep{1992ApJ...396..469H, 1995A&A...295...47Q, 1997NewA....2..449G}
would explain such fluctuations as long as the emitting region has a small enough angular size.

Suppose that the first two data points following FRB 150418 are indeed the afterglow of the FRB, the existence of a bright radio
host is still puzzling. One may consider the following possibilities. 

(1) First, the host may be related to the star formation in the host.
Assuming that the quiescent radio emission of the host originates from star formation,
the star formation rate (SFR) estimated with radio emission is $10^{1.7} {\rm M_{\sun}~yr^{-1}}$
\citep{2012ARA&A..50..531K}, two orders of magnitude larger than the upper limit
estimated from H$\alpha$ emission line \citep{2016Natur.530..453K}.
It indicates an inconsistency between the emission line-estimated SFR and the
continuum-estimated SFR.
Such a discrepancy also occasionally seen in SGRB hosts. For example,
the host of GRB 050509B, which is an elliptical galaxy, also shows no SFR by using emission lines as an indicator,
$< 0.1\ \rm M_{\sun}\ yr^{-1}$ \citep{2009ApJ...690..231B},
while the SFR estimated by UV emission is
$16.9\ \rm M_{\sun}\ yr^{-1}$\citep{2009ApJ...691..182S}.
Although such a possibility is not ruled out,
it is not favored.
Within this picture, the host flux is not expected to fluctuate significantly. 
If the VLA rebrightening 
reported by \cite{2016ATel.8752....1W} is not due to a mis-calibration between VLA and ATCA, 
it strongly disfavors this possibility.

(2) Second, the quiescent radio flux is from an underlying low-activity AGN, while the FRB is related to a compact-star-merger
event \citep{2016arXiv160208086Z} within the host galaxy of the AGN. This is not impossible, since the host galaxy appears as an elliptical galaxy
with little star formation, which is consistent with being a host of compact star mergers 
\citep{2005Natur.437..851G, 2005Natur.438..994B, 2014ARA&A..52...43B}.
The probability of having such a weak AGN host may be low, and is worth investigating. 
As an elliptical galaxy with a stellar mass 10$^{11}$ $M_{\sun}$ \citep{2016Natur.530..453K}, 
the center black hole mass of the putative FRB host 
may be estimated using the $M_{\rm BH}-M_{\rm *}$ relation
\citep{2004ApJ...604L..89H}, which gives $M_{\rm BH}=10^{8.2} M_{\rm \sun}$.
If the radio quiescent emission is trully from the central black hole,
It indicates a radio luminosity $4.2\times10^{39}\ \rm erg/s$.
The X-ray luminosity anticipated from the black hole activity fundamental plane
is 1.4 $\times 10^{43}$ erg/s, which is smaller than 
the X-ray upper limit from Swift \citep{2016Natur.530..453K}.
Although the optical spectrum of the host disfavors an AGN at the center,
it is supported by the consistency between its radio spectrum and
those of AGNs
\citep{2016arXiv160304421V}.
It may be an low luminosity AGN
or radio analog to the X-ray bright Optical galaxy (XBONG)
\citep{2004ApJ...612..724Y, 2003MNRAS.344L..59M}. 
Deep X-ray monitoring might be the best way to investigate the AGN possibility, and
the origin of the radio quiescent emission of the host.
In any case, the existence of an AGN does not rule out 
the possibility that the first two data points are due to the FRB 150418 afterglow.

(3) The third probability is that both afterglow
and AGN are true (similar to the second possibility), 
but the FRB may be related to the AGN itself. 
However, no viable FRB model
has been proposed to be produced in an AGN environment. One difficulty would be the time scale. Whereas the shortest time scale
for a supermassive BH may be hours, the typical FRB duration is at most milliseconds. More work is needed to explore this possibility.

\acknowledgments
We thank helpful communications and comments from Edo Berger, Evan Keane
and Weimin Yuan.
This work is partially supported by NASA NNX15AK85G and NNX14AF85G.

\end{document}